\documentclass[prl,twocolumn,showpacs,superscriptaddress,longbibliography]{revtex4-2}

\usepackage{amsmath,amssymb}
\usepackage{xcolor}
\usepackage{graphicx}
\usepackage{braket}
\usepackage[utf8]{inputenc}
\usepackage{bm}

\begin{document}
\date{\today}
\title{Scalable generation of multi-photon entangled states by active feed-forward and multiplexing}
\author{Evan Meyer-Scott}
\affiliation{Paderborn University, Integrated Quantum Optics, Institute for Photonic Quantum Systems (PhoQS), 33098 Paderborn, Germany}
\author{Nidhin Prasannan}
\affiliation{Paderborn University, Integrated Quantum Optics, Institute for Photonic Quantum Systems (PhoQS), 33098 Paderborn, Germany}
\author{Ish Dhand}
\affiliation{Institut für Theoretische Physik and Center for Integrated Quantum Science and Technology (IQST), University of Ulm, 89069 Ulm, Germany}
\author{Christof Eigner}
\affiliation{Paderborn University, Integrated Quantum Optics, Institute for Photonic Quantum Systems (PhoQS), 33098 Paderborn, Germany}
\author{Viktor Quiring}
\affiliation{Paderborn University, Integrated Quantum Optics, Institute for Photonic Quantum Systems (PhoQS), 33098 Paderborn, Germany}
\author{Sonja Barkhofen}
\email{sonja.barkhofen@uni-paderborn.de}
\affiliation{Paderborn University, Integrated Quantum Optics, Institute for Photonic Quantum Systems (PhoQS), 33098 Paderborn, Germany}
\author{Benjamin Brecht}
\affiliation{Paderborn University, Integrated Quantum Optics, Institute for Photonic Quantum Systems (PhoQS), 33098 Paderborn, Germany}
\author{Martin B. Plenio}
\affiliation{Institut für Theoretische Physik and Center for Integrated Quantum Science and Technology (IQST), University of Ulm, 89069 Ulm, Germany}
\author{Christine Silberhorn}
\affiliation{Paderborn University, Integrated Quantum Optics, Institute for Photonic Quantum Systems (PhoQS), 33098 Paderborn, Germany}

\begin{abstract}
Multi-photon entangled quantum states are key to advancing quantum technologies such as multi-party quantum communications, quantum sensing or quantum computation. Their scalable generation, however, remains an experimental challenge. Current methods for generating these states rely on stitching together photons from probabilistic sources, and state generation rates drop exponentially in the number of photons. Here, we implement a system based on active feed-forward and multiplexing that addresses this challenge. We demonstrate the scalable generation of four-photon and six-photon GHZ states, increasing generation rates by factors of 9 and 35, respectively. This is consistent with the exponential enhancement compared to the standard non-multiplexed approach that is predicted by our theory. These results facilitate the realization of practical multi-photon protocols for photonic quantum technologies.
\end{abstract}

\pacs{
03.67.-a,
42.50.-p,
 03.67.Bg,
 42.50.Dv, 
 42.50.Ex 
}

\maketitle

\emph{Introduction.---}
Entangling multi-partite quantum systems is an important task in quantum technologies \cite{pan2012multiphoton}.
Large entangled photonic states are key to multi-party communication \cite{su2016quantum, jin2006three,hao2001controlled}, where they provide distinct advantages over pairwise entanglement such as increased secret key rates and robustness to some types of noise \cite{malik2016multi,epping2017multi,hu2020experimental,proietti2021experimental}.
Large entangled states are also critical for obtaining a quantum advantage in cluster-state linear optical quantum computation \cite{raussendorf2003measurement,walther2005experimental,prevedel2007high,gao2011experimental,yao2012experimental,adcock2019programmable} and quantum-enhanced metrology \cite{dowling2008quantum}. 
The scalable generation of such states especially at telecommunications wavelengths remains, however, an outstanding challenge.
One prominent example of multi-photon entangled states are Greenberger-Horne-Zeilinger (GHZ) states \cite{Greenberger1990Bell}. 
They are typically generated via postselection by interfering photons from polarization Bell pairs – photon pairs that are maximally entangled in their polarization degree of freedom – at passive linear optical elements \cite{bouwmeester1999observation, pan2001experimental,lu2007experimental,megidish2012resource, takeda2019demand,wu2021robust,xu2022experimental}.
For a 4 photon GHZ state two polarization Bell pairs must be generated e.g. in a state like 
\begin{equation} \nonumber
|\Psi\rangle = \frac{1}{\sqrt{2}}(|H_1 H_2\rangle +|V_1 V_2\rangle)\otimes \frac{1}{\sqrt{2}}(|H_3 H_4\rangle +|V_3 V_4\rangle)
\end{equation}
where $H_i$ and $V_i$ denote horizontal and vertical polarized photons in path $i$.
In a next step paths 2 and 4 are fused at a polarising beam splitter (PBS) and, after post-selecting events with one photon exiting at each output of the PBS, the desired GHZ state
\begin{equation} \nonumber
|\Psi_\mathrm{GHZ}^{(4)}\rangle = \frac{1}{\sqrt{2}}(|H_1 H_2 H_3 H_4\rangle +|V_1 V_2 V_3 V_4\rangle)
\end{equation}
is formed. 
The operation implicitly assumes that the photons in the combined paths are indistinguishable in all degrees of freedom to guarantee their reliable interference.
This scheme directly generalises to $N$ Bell pairs forming a post-selected $2N$-photon GHZ state. 
The prevailing way of generating the required Bell pairs is to use probabilistic photon-pair sources. 
Each source exhibits a very low generation probability $p\ll 1$ (typical values are on the order of few percent) and the total probability for generating $N$ Bell pairs in $M$ trials is given by $1-(1-p^N)^M \approx Mp^N$.
Thus, the generation probability decreases exponentially in the number of pairs $N$ and grows linearly with $M$, the number of trials. 
The current state of the art is the production of post-selected 12-photon GHZ states, which are detected with a rate of one per hour despite each source generating Bell pairs at a rate of 2 MHz \cite{zhong201812}.  
\begin{figure*}[t]
	\includegraphics[width=0.8\textwidth]{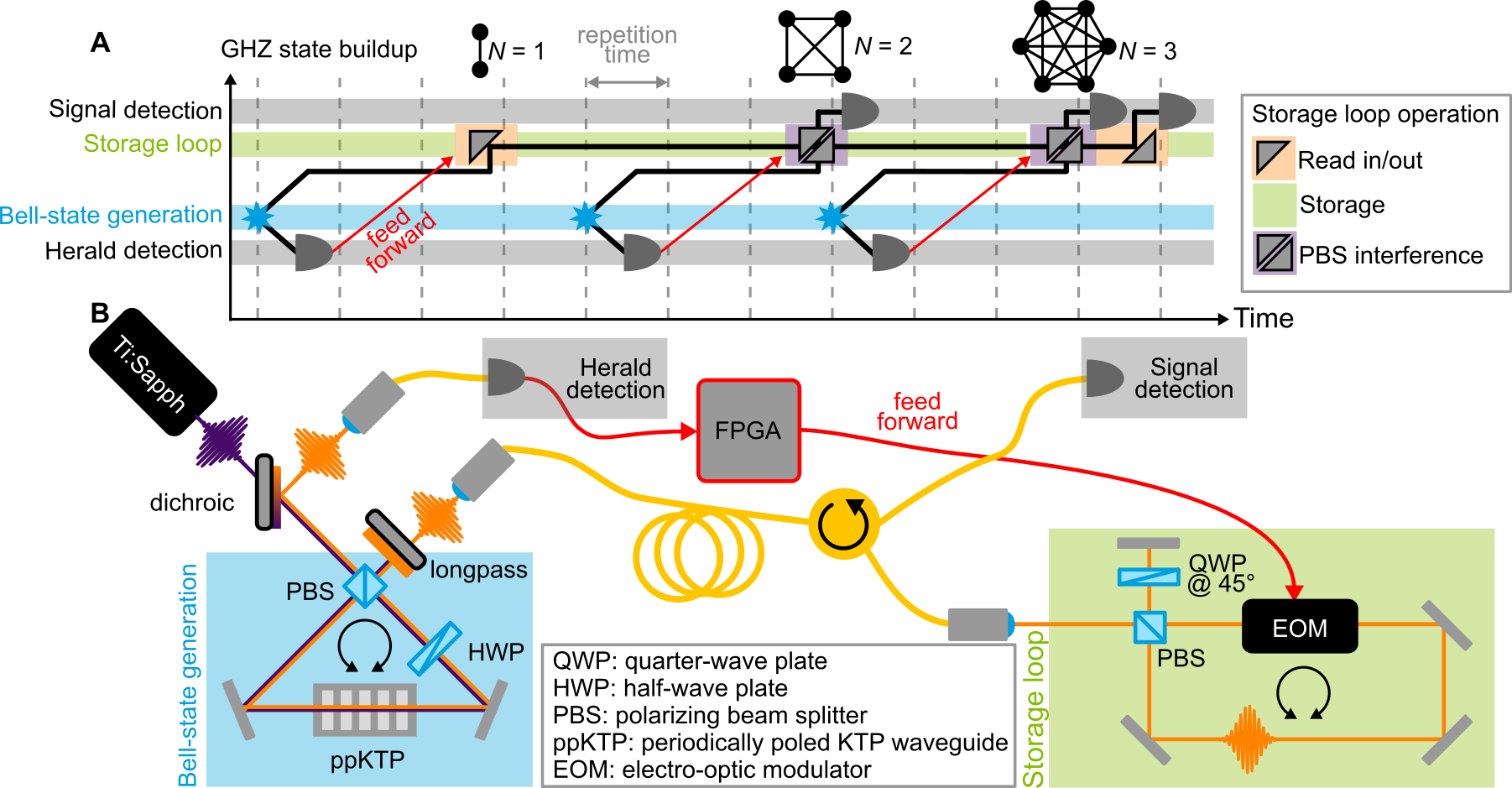}
	\caption{\textbf{(A)} Operating principle of our approach. Bell pairs are generated sequentially. The detection of one photon triggers the feed-forward including a field programmable gate array (FPGA), which in turn controls the operation mode of an all-optical storage loop. Possible operation modes are ``Read in/out'' (orange), ``Storage'' (green), or ``PBS interference'' (purple), selected by an appropriate switching of the electro-optic modulator (EOM). $2N$-fold coincidences confirm the build-up of $2N$-photon GHZ states. \textbf{(B)} Sketch of the experimental setup. A Ti:Sapph laser with a wavelength of 775\,nm pumps a polarization Bell-state source based on parametric down-conversion in a Sagnac configuration (blue area). One photon of each emitted Bell pair is detected and triggers the feed forward (red arrows), the other photon is sent to our all-optical storage loop (green area), where it is stored until it is brought to interference with the subsequent qubit. For more information, please see the text.}
	\label{fig:setup}
\end{figure*}
\\
One strategy to enhance the generation rates is the utilization of multiplexing schemes with feed-forward.
This has been proposed originally for single-photon generation \cite{migdall2002tailoring,pittman2002single} and has later been extended to the simultaneous generation of multiple single photons \cite{nunn2013enhancing, gimeno2017relative}. 
The concept, however, also offers  a considerable potential when applied to polarization qubits, which, in turn, allow for the generation of multi-photon entangled states \cite{dhand2018proposal}. 
In contrast to spatial multiplexing schemes, time multiplexing keeps the required number of experimental components constant and, thus, constitutes the most resource efficient approach \cite{meyer2020single}.
In general, the underlying idea of time multiplexing for enhancing state generation rates is to repeat a probabilistic operation, e.g., photon pair or Bell state creation, $M$ times and store a successful outcome (or parts of it) in a quantum memory until the subsequent operation succeeds and all photons are actively coupled out. 
For $N$ time-multiplexed probabilistic sources the probability for generating $N$ successful generation events in $M$ trials grows as $(1-(1-p)^M)^N \approx (Mp)^N$, providing an exponentially increasing prefactor $M^{(N-1)}$ in stark contrast with the non-multiplexed case \cite{kaneda2017quantum}.

In this work, we demonstrate the enhanced generation of multi-photon GHZ states using time multiplexing and active feed-forward. 
We combine an integrated-optics Bell state source \cite{meyer2018high}, that emits Bell pairs at telecom wavelengths, with a high-efficiency, high-bandwidth all-optical polarization qubit memory. 
The latter not only stores the polarization qubit until the next Bell pair is successfully generated, but it additionally acts as polarizing beam splitter for the GHZ state creation, when it is triggered by the next heralding event in an active feed-forward operation.

\emph{Operating Principle and Experimental Setup.---}
In the following, we describe the operating principle and  the experimental setup  of the feed-forward, see Figure~\ref{fig:setup}A. 
A Bell state source (blue bar) is pumped with a train of pump pulses (grey dashed lines) which defines the number of trials $M$. 
Upon a successful pair-generation event (blue star), we detect one of the photons as a herald. 
This detection triggers a feed-forward signal (red lines) that actively controls the operation mode of an all-optical, polarization insensitive storage loop.
The second photon is delayed in an optical fibre to account for the latency in the feed-forward, and read into the storage loop (orange square) and then stored (green bar) until the next Bell state is generated. 
Upon detection of the second herald photon, the storage loop is set to act as a PBS (purple square), such that the newly generated photon and the previously stored photon interfere. 
The outcome of this interference is probabilistic and in 50\% of the cases one photon remains inside the storage loop, while the other photon is detected by the signal detector. 
This time-multiplexed operation repeats until the desired state size (in the case of Figure~\ref{fig:setup}A a six-photon GHZ state generated from three Bell states) is achieved, upon which the remaining photon inside the storage loop is actively read out (orange square) and detected.
Successful state generation is achieved in those cases where each detector clicks once, an equivalent post-selection as in the non-multiplexed setting. 

\begin{figure*}[t]
	\includegraphics[width=0.8\textwidth]{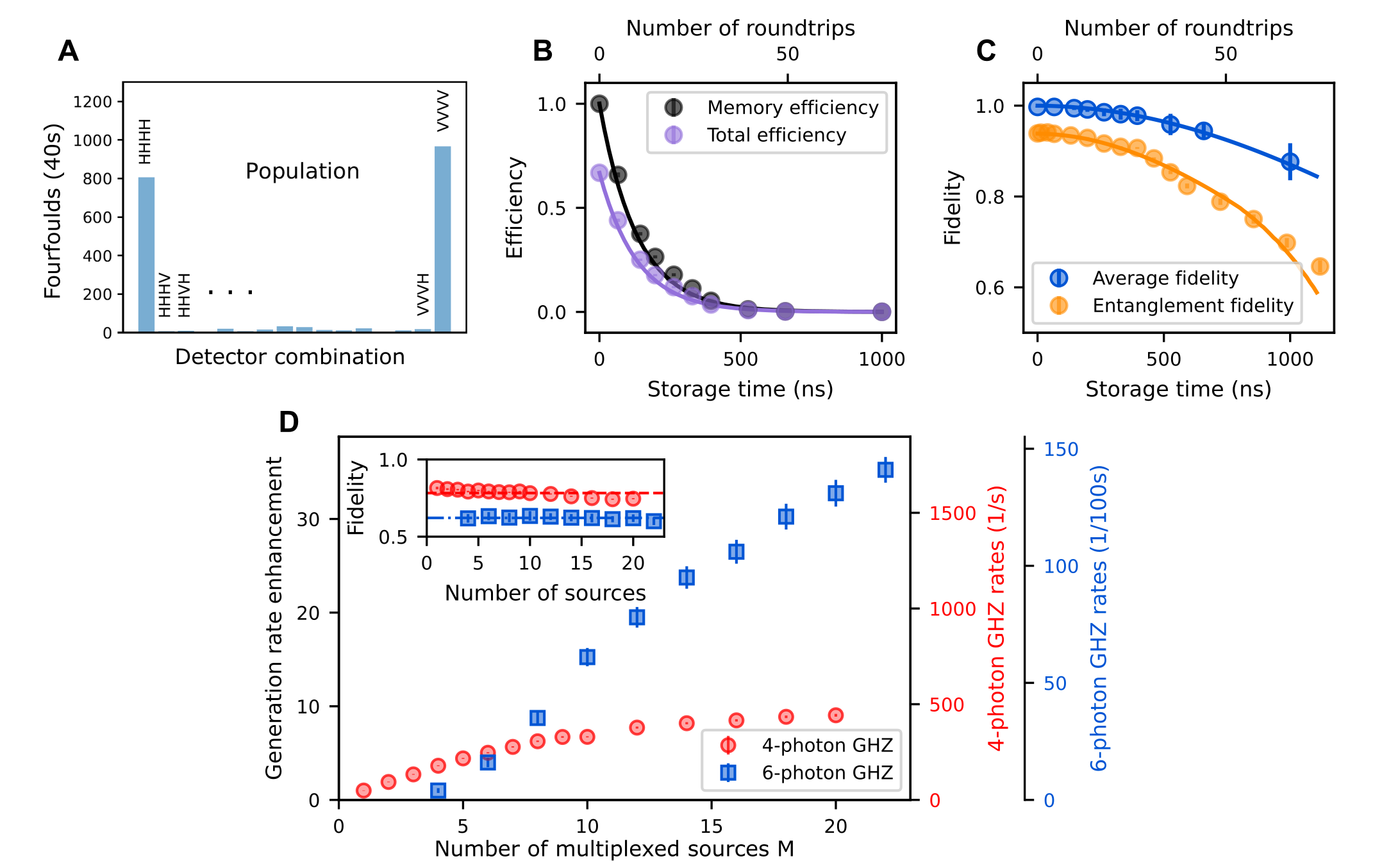}
	\caption{\textbf{A} Raw coincidence count data for 4-photon GHZ state for $M = 2$ sources (see supplementary figure S8 for the 6 photon data) \textbf{B}  Total and storage efficiency of the all-optical storage loop. We obtain a storage efficiency of 91\% with a corresponding lifetime of 131 ns. \textbf{C} Average polarization single-qubit fidelity that is, fidelity of a retrieved polarization qubit with respect to its initial state averaged over many input polarizations, and entangled-pair fidelity after storage. For up to 20 round trips in the storage loop, the fidelities stay roughly constant. We note that the entanglement fidelity is limited by the performance of the Bell-state source. \textbf{D} Generation rate enhancement for four- and six-photon GHZ states as a function of the number of successive uses of the Bell-state source $M$. The observed saturation is due to the limited efficiency of the storage loop. We find an enhancement of up to factor 35 for six-photon GHZ states. The inset displays the state fidelities, which stay constant regardless of the number of multiplexed sources. For more information see the text.}
	\label{fig:results}
\end{figure*}

In Figure~\ref{fig:setup}B, we sketch our experimental implementation of the scheme of Figure~\ref{fig:setup}A. 
Our Bell state source is a periodically poled potassium titanyl phosphate waveguide source in a Sagnac configuration \cite{meyer2020single}, which is pumped with pulses from a titanium sapphire oscillator with a repetition rate of 76\,MHz and a central wavelength of 775\,nm. 
It generates Bell states $|\Psi\rangle =  \frac{1}{\sqrt{2}}(|H_1V_2\rangle +|V_1H_2\rangle)$ with a central wavelength of 1550\,nm and a fidelity of up to 96\%. 
We perform a polarization-resolved detection of  photon 1 – the herald – with an overall efficiency of 43\% \cite{klyshko1980use}. 
This detection triggers a field programmable gate array (FPGA), which generates the electric feed-forward signal that sets the operation mode of the storage loop (for more information on electronics and synchronization see Supplementary Material S5). 
The second photon – the signal – is coupled to a 300\,m long single mode fibre to compensate for the electric latency of the feed forward and allow all herald photons to be detected for all possible time-bin combinations before signal photons enter the loop (roughly 700\,ns). 
After the fibre, the second photon is coupled into our all-optical storage loop, thereby its polarization is swapped  via the double pass through a quarter-wave plate (QWP) at $45^\circ$ in the retroreflective delay line. 
The optical storage loop is synchronized to the repetition rate of the laser (76\,MHz, corresponding to 13.16\,ns separation between successive pulses) and set up such that the two polarization components of the signal photon travel in opposite directions. 
We place a fast, electro-optic modulator (EOM) at an asymmetric position next to the input/output polarizing beam splitter of the loop.
Its fast rise- and fall-time of around 5\,ns allows for the selected polarization conversion of each of the clockwise and counterclockwise travelling components individually.
This facilitates in particular the operation mode ``PBS interference'', where the storage loop acts as polarizing beam splitter for an input photon and a stored photon (see the Supplementary Material S1 for a detailed description of the operations modes and the corresponding EOM switchings). 
As described in Figure~\ref{fig:setup}A, the signal photon is stored in the storage loop until the next Bell state is generated. 
Then, the storage loop is set to ``PBS interference'' mode and a four-photon GHZ state can be created, given that one of the signal photons leaves the loop and is detected by our polarization-resolving signal detector. 
This procedure can be repeated until the desired state size is achieved.
Finally, the storage loop is set to ``Read in/out'' mode and the last photon inside the loop is coupled out of the loop and sent to the detection stage.
For previously stored photons, if there was no other successful pair generation event within the next $M$ trials, i.e. pump pulses, the stored photon is released and the run discarded.
The number $M$ thus determines the maximum waiting time in the storage, but at the same time the number of time-multiplexed sources involved in the GHZ state generation process.

\emph{Results.---}
\begin{figure*}[t]
	\includegraphics[width=0.8\textwidth]{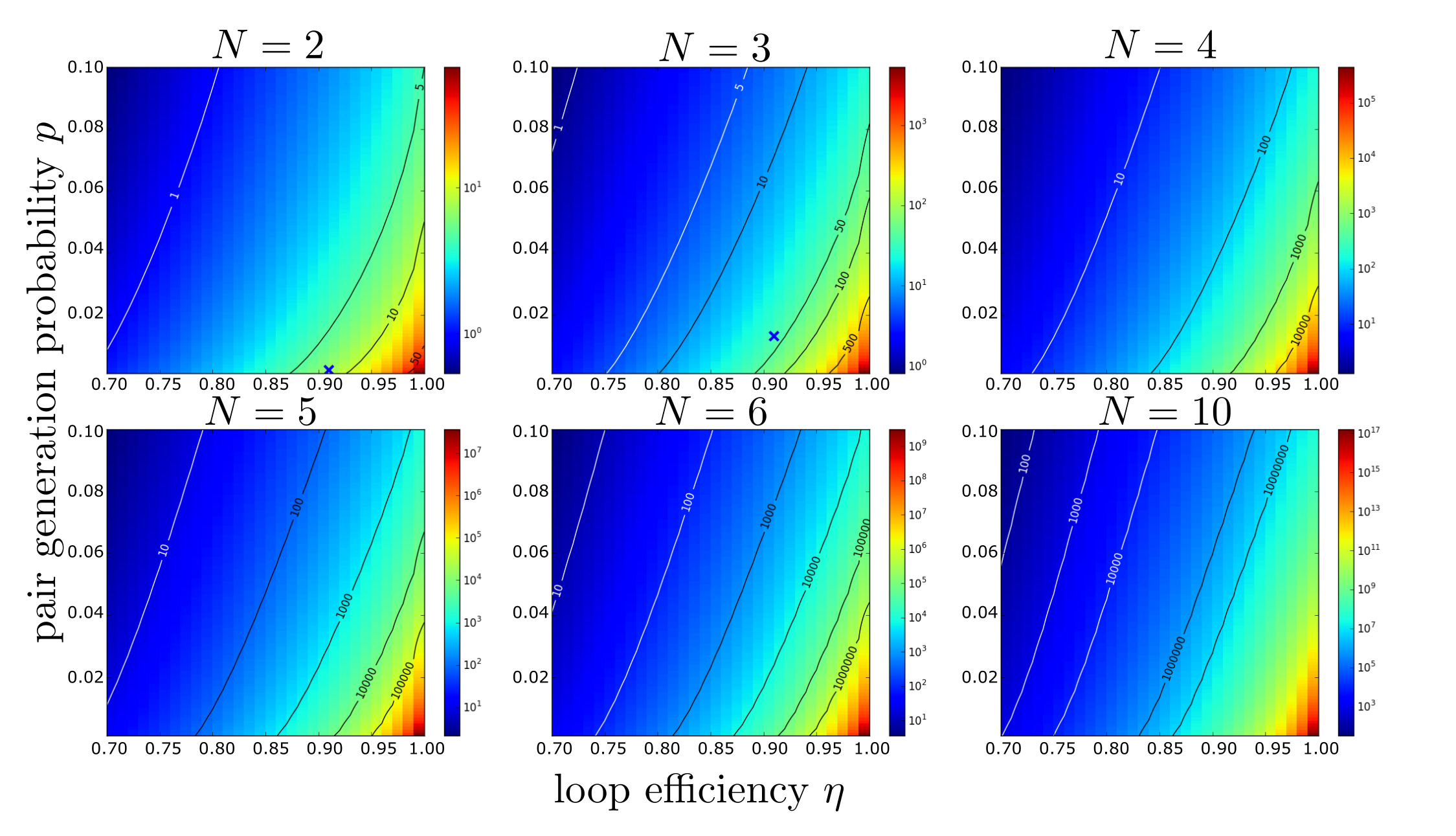}
	\caption{Calculated rate enhancement as a function of the Bell-pair production probability and the memory efficiency for $N=2,3,4,5,6,10$ Bell pairs and optimized $M$ parameter. Our measurements for $N = 2$ and $N = 3$ (blue crosses) are well described by the theory (in the experiment we used up to $M = 22$ time-multiplexed sources). Larger states are predicted by the same theory to experience stronger enhancement, up to nine orders of magnitude for $N=6$ that is, 12-photon GHZ states, the actual state of the art. For more information on the expected absolute rates see Supplemental Material S4.}
	\label{fig:simulation}
\end{figure*}
Next, we present the experimental results (Fig.~\ref{fig:results}) and discuss the potential of the presented time-multiplexing approach for generating large states. 
In subfigure~\ref{fig:results}A we present the measured 4-photon counts in the $H/V$ basis (see Supplementary Material S6 for more information on population and coherence for 4- and 6-photon GHZ state).
Figures~\ref{fig:results}B and ~\ref{fig:results}C are a performance study of our storage loop. 
Where present, solid curves are the result of our numerical model of the system, including losses and misalignments.
From the measured total efficiency as function of storage time (purple data points), we extract a memory efficiency of 91\% and a $1/e$ – lifetime of 131\,ns, which corresponds to around 11 round trips.
More importantly, both the average fidelity (blue data points) of a stored polarization qubit and the entanglement fidelity (orange data points) of a Bell state decrease only slowly during storage. 
Especially the region up to 20 round trips, i.e. 20 multiplexed sources,  – the relevant region for this work – shows hardly any decay with the average fidelity decreasing only from 99.7\% to 98.5\%. 
We note that these measurements were taken for photons with a central wavelength of 1550\,nm and a bandwidth of 0.52\,THz, setting our storage loop apart from solid-state based quantum memories, which typically operate at shorter wavelengths and narrower bandwidths (see the Supplemental Material S1 A). 
Finally, Figure~\ref{fig:results}D shows the measured four- (red) and six-photon (blue) GHZ state rates as a function of the number $M$ of successive uses of our Bell-state source.
Our measurements demonstrate an increase of a factor of 9.7 for the preparation probability of four-photon GHZ states and of 35 for six-photon GHZ states.
This yields net detection rates of around 450 four-photon states per second and 1.4 six-photon states per second (note the different units of the Y-axes). 
The inset shows the measured state fidelities, which were extracted by measuring the population in the H/V basis and the respective coherences following \cite{wang2016experimental}. 
We measure state fidelities of 81\% for the four-photon GHZ state and 63\% for the six-photon GHZ state, both clearly above the classical bound of 50\% and limited by the initial fidelity of the underlying Bell states. 
We also see from the fidelity measurements that multiplexing does not degrade the state fidelity.
In summary, our results demonstrate that the generation probability of GHZ states can be increased exponentially in the number of photon pairs while basically maintaining the targeted high state fidelities.

Finally, we discuss the potential of our setup for achieving even larger GHZ states.
Figure~\ref{fig:simulation} shows the expected GHZ rate increase (color scale) as a function of the memory loop efficiency and the pair-generation probability per source for $N=2\dots10$ Bell pairs. 
For $N=1$ there is no rate increase because we only introduce additional losses with an imperfect memory. 
For $N>1$, we find large regions of enhanced generation rates for reasonable experimental parameters. 
We see that our measured rate increases (blue crosses) agree well with the theory, and with the same theory we predict a rate increase by more than nine orders of magnitude over the state of the art for 12-photon GHZ states for realistic experimental settings. 
In fact, using the pair production probability, detection and heralding efficiencies from \cite{zhong201812}, we predict a detection rate of around one 12-photon GHZ state per second using our system which compares favourably to the 1 per hour detection in \cite{zhong201812} (see Supplementary Material S4).

As an additional use of this platform, the basic principle of the storage loop can be extended directly to generate linear cluster states and post-selected CPHASE gates by including a second EOM in the loop or a fast attenuator, respectively (Supplementary material S3), signifying the broad applicability of our system for complex quantum state generation.

\emph{Conclusion.---}
In conclusion, we have built a system based on time multiplexing and active feed-forward that provides an exponential enhancement of $M^{N-1}$ in the generation rates of multi-photon entangled quantum states. 
We have experimentally demonstrated the generation of four- and six-photon GHZ states and saw a rate increase of up to a factor of 35 compared to non-multiplexed scenarios when multiplexing up to 22 sources confirming our theoretical predictions. 
In addition, the state fidelity does not suffer, meaning that we generate the same high-quality states at higher rates.
Finally, based on our theory we demonstrated that the generation rate increase grows with the state size, resulting in a potential three-order of magnitude rate increase for the generation of 12-photon GHZ states – the current state of the art. 
Our system brings into reach applications that exploit the power of large, entangled states, as it enables measurement times that are short enough to be of interest to a wide range of users.

\begin{acknowledgments}
This work was supported by the ERC project QuPoPCoRN (Grant no. 725366), the ERC Synergy grants BioQ (Grant no. 319130) and HyperQ (Grant no. 856432), the Alexander-von-Humboldt Foundation and the EU project AsteriQs (Grant no. 820394).
\end{acknowledgments}

%

\end{document}